\documentclass[pra,aps,twocolumn,preprintnumbers,showpacs,superscriptaddress]{revtex4-1}
\usepackage{graphics,graphicx,epsfig,bm,amsmath,amsthm,amssymb}

\usepackage{lipsum} 

\usepackage{amsmath}
\usepackage{subfigure}

\usepackage{bm}
\usepackage{bbm}
\usepackage{upgreek}
\usepackage{multirow}
\usepackage{natbib}
\usepackage{color}
\usepackage[usenames,dvipsnames]{xcolor}
\usepackage[version=3]{mhchem} 
\usepackage[T1]{fontenc}       
\usepackage[export]{adjustbox} 

\usepackage[a4paper,colorlinks=true,
linkcolor=blue,citecolor=blue,
pdfauthor={ },
pdftitle={ },
pdfsubject={ },
pdfkeywords={ }]{hyperref}

\makeatother

\begin{document}
	
	\title{Nanoscale electrometry based on a magnetic-field-resistant spin sensor}
	
	\author{Rui Li}
	\altaffiliation{These authors contributed equally to this work.}
	\author{Fei Kong}
	\altaffiliation{These authors contributed equally to this work.}
	\author{Pengju Zhao}
	\author{Zhi Cheng}
	\author{Zhuoyang Qin}
	\author{Mengqi Wang}
	\author{Qi Zhang}
	
	\author{Pengfei Wang}
	\author{Ya Wang}
	\email{ywustc@ustc.edu.cn}
	\author{Fazhan Shi}
	\email{fzshi@ustc.edu.cn}
	
	\author{Jiangfeng Du}
	\email{djf@ustc.edu.cn}
	\affiliation{Hefei National Laboratory for Physical Sciences at the Microscale and Department of Modern Physics, \\
		University  of  Science  and  Technology of  China (USTC), Hefei, 230026, China}
	\affiliation{CAS Key Laboratory of Microscale Magnetic Resonance, USTC, Hefei, 230026, China}
	\affiliation{Synergetic Innovation Center of Quantum Information and Quantum Physics,\\ USTC, Hefei, 230026, China}


	\begin{abstract}
		The nitrogen-vacancy (NV) center is a potential atomic-scale spin sensor for electric field sensing. However, its natural susceptibility to the magnetic field hinders effective detection of the electric field. Here we propose a robust electrometric method utilizing continuous dynamic decoupling (CDD) technique. During the CDD period, the NV center evolves in a dressed-state space, where the sensor is resistant to magnetic fields but remains sensitive to electric fields. As an example, we use this method to isolate the electric noise from a complex electro-magnetical environment near diamond surface via measuring the dephasing rate between dressed states. By reducing the surface electric noise with different covered liquids, we observe an unambiguous relation between the dephasing rate and the dielectric permittivity of the liquid, which enables a quantitative investigation of electric noise model near diamond surface.
	\end{abstract} 
	\maketitle

	Characterization of electrical properties and comprehension of the dynamics in nanoscale become significant in the development of modern electronic devices, such as semiconductor transistors and quantum chips, especially when the feature size has shrunk to several nanometers. For various application scenarios, a rich toolbox of nanoscale electrometry has been established. Scanning probe-based techniques, such as single-electron transistor (SET) \cite{yoo1997scanning} and single-electron electrostatic force microscopy (SE-EFM) \cite{schonenberger1990observation}, have the ability of imaging of single charge on the surface with high spatial resolution. X-ray methods \cite{sakdinawat2010nanoscale, holler2017high} and in situ atomic probe \cite{zhang2019single} can provide internal information of electronic devices. Recently, the nitrogen-vacancy (NV) center in diamond, an atomic-scale spin sensor, has shown to be an attractive electrometer \cite{dolde2011electric, dolde2014nanoscale, iwasaki2017direct, mittiga2018imaging, broadway2018spatial}. Benefited from the in situ compatibility with diamond-based semiconductor devices \cite{iwasaki2017direct, broadway2018spatial} and the potential of electric-field imaging by combing the scanning technology \cite{maletinsky2012robust}, electrometry by using the NV center would advantage various sensing and imaging applications.
	
	As a spin sensor, the NV center is naturally sensitive to the magnetic field due to the Zeeman effect. Actually, it has already been demonstrated as an outstanding probe for nanoscale sensing and imaging of magnetic field \cite{maze2008nanoscale, balasubramanian2008nanoscale, sushkov2014magnetic, grinolds2014subnanometre, shi2015single}. However, the Stark effect in the NV ground state is usually a negligible perturbation, because it is much weaker than the Zeeman effect \cite{dolde2011electric}. So the key of NV center-based electrometry is carefully suppressing the influence of magnetic fields \cite{dolde2011electric, dolde2014nanoscale, iwasaki2017direct, broadway2018spatial, mittiga2018imaging}. The static component of the magnetic field is usually canceled by sophisticated alignment of the external magnetic field \cite{dolde2011electric, iwasaki2017direct}, while the dynamic component (i.e., magnetic noise) induced by bath spins is previously suppressed by a strong non-axial magnetic field \cite{dolde2014nanoscale, broadway2018spatial}, although with compromise of sensitivity. For example, the non-axial magnetic field will amplify the transverse hyperfine coupling with the adjacent nitrogen nuclear spin \cite{dolde2014nanoscale}, and thus an obvious energy splitting will appear with reduced signal contrast. For an electric field sensing, some magnetic-dipole-forbidden transitions such as double-quantum transition can also be used to selectively detect electric fields, but only with high ($\geqslant \ $MHz) frequencies \cite{mamin2014multipulse, myers2017double}.
	
	Here, we demonstrate a novel NV center based electrometry combined with the continuous dynamical decoupling (CDD) technique \cite{fanchini2007continuously, chaudhry2012decoherence, cai2012robust, xu2012coherence}, where the continuous driving fields provide a magnetic-field-resistant dressed-state space. In this space, the Stark effect is preserved, and then we observe a simple linear dependence of the energy levels on the magnitude of electric fields. Moreover, we use this method to unambiguously investigate the electric noise environment near the diamond surface, where the magnetic noise is strong and ubiquitous \cite{ romach2015spectroscopy, rosskopf2014investigation}. We measure the purely electric noise induced dephasing between dressed states of near-surface NV centers. Since the surface electric noise can be reduced with covered liquid \cite{kim2015decoherence}, we can obtain a quantitative relation between the dephasing rates and the dielectric permittivities of covered liquids, and then gain insight into the electric noise model near diamond surface.
	
	\begin{figure}
		\centering
		\includegraphics[scale=0.66]{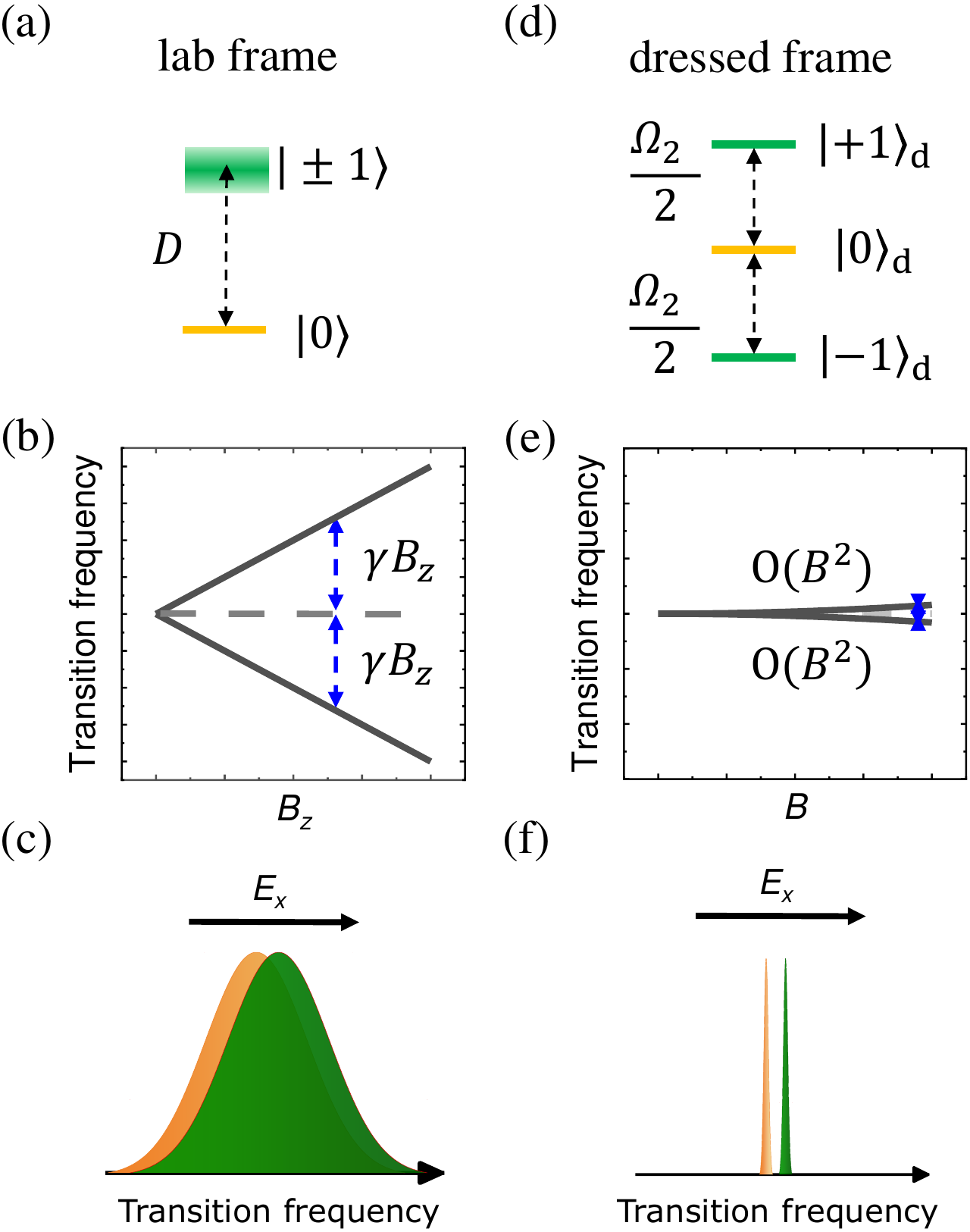}
		\caption{\textbf{Basic concept of NV center-base electrometer.} \textbf{(a)} The energy structure of the NV center in the lab frame (ground states, $m_s=0$ ($\vert 0\rangle$ henceforth) and degenerate $m_s=\pm 1$ ($\vert \pm 1\rangle$ henceforth)). \textbf{(b)} The transition frequencies between the bare states in the lab frame are linear with the axial magnetic field. \textbf{(c)} The natural broadening of the transition frequency is dominated by the magnetic field noise. \textbf{(d)} The energy structure of the NV center in the dressed frame by continuous microwave driving. \textbf{(e)} The transition frequencies between the dressed states have zero first-order dependence on the magnetic field. \textbf{(f)} The magnetic-noise-induced broadening is strongly suppressed, and thus the electric-field-induced frequency shift can be easily resolved.}
		\label{Fig-1}
	\end{figure}
	
	NV center is a defect in diamond, which consists of a substitutional nitrogen and an adjacent vacancy \cite{nizovtsev2003nv}. The outstanding achievements of the NV center benefit from its remarkable properties such as, most notably, convenient state polarization and readout by a 532-nm laser \cite{steiner2010universal} and long coherence time due to the spin-purity environment \cite{knowles2014observing}. The Hamiltonian of the NV center in the lab frame is given by \cite{mittiga2018imaging, van1990electric}:
	\begin{equation}\label{H0}
		\begin{split}
		H_0&=(D+d_\parallel E_z)S_z^2 +\gamma \vec{B}\cdot\vec{S}\\
		&- d_\perp E_x (S_x^2-S_y^2)+d_\perp E_y (S_x S_y+S_y S_x),
		\end{split}
	\end{equation} 
	where $D/h=2.87$ GHz is the zero-field splitting exhibited by the NV's ground states, $d_\parallel/h  = 0.35 \pm 0.02\ \rm{Hz\ cm\ V^{-1}}$ and $d_\perp/h = 17 \pm 3\ \rm{Hz\ cm\ V^{-1}} $ are the axial and non-axial electric dipole moments, $\gamma = 28.03\ \rm{ GHz/T}$ is the gyromagnetic ratio of the electron spin, $\vec{S}$ is the spin operator, $\vec{B}$ and $\vec{E}$ are the magnetic and electric field, respectively. The energy structure in the lab frame is shown in Fig.~\ref{Fig-1}(a). 
	Generally, the zero-field splitting is the dominant item, and thus only the axial magnetic field $B_z$ matters. The transition frequencies from $\vert 0\rangle$ to $\vert \pm 1\rangle$ are proportional to the axial magnetic field $B_z$, which is known as the Zeeman effect (Fig.~\ref{Fig-1}(b)). The presence of $B_z$ will also suppress the non-axial Stark term which does not commute with $S_z$. Though the axial Stark term is unaffected, the detection of $E_z$ is much inefficient because $d_\parallel$ is 50 times smaller than $d_\perp$. Furthermore, the ubiquitous magnetic noise from surrounding baths, especially for near-surface NV centers, will  broaden the transition frequency (Fig.~\ref{Fig-1}(c)), which hinders the observation of energy level shift induced by electric fields.
	
	For the purpose of electric field sensing, both the preservation of the non-axial Stark term and the suppression of magnetic noise are important. In this work, we show that these two issues can be simultaneously addressed by introducing a continuous driving field:
	\begin{equation}\label{H}
	H_1=\Omega_1 \cos(D t+\frac{2\Omega_2}{\Omega_1}\sin(\Omega_1 t))S_x,
	\end{equation}
	where $\Omega_1$ is the amplitude of the continuous microwave and $\Omega_2$ controls the phase modulation which is used to stabilize the microwave amplitude \cite{cohen2017continuous}. 
	The Hamiltonian in the dressed frame is given as \cite{SOM}:
	\begin{equation}\label{Heff}
	H_{\rm d}=\frac{\Omega_2}{2}S_{{\rm d},z} + \frac{d_\parallel E_z+3d_\perp E_x}{2}S_{{\rm d},z}^2+O(B^2),
	\end{equation}
	here the subscript "d" here indicates the operators in the dressed frame. The energy structure in the dressed frame is shown in Fig.~\ref{Fig-1}(d), and the corresponding eigenstates are:
	\begin{equation}\label{ES}
		\begin{split}
			\vert+1\rangle_{\rm d} &= \frac1{\sqrt{2}}(\vert+1\rangle +\vert-1\rangle),  \\       
			\vert0\rangle_{\rm d} &= \frac1{\sqrt{2}}(\vert+1\rangle -\vert-1\rangle),\\                   
			\vert-1\rangle_{\rm d} &= \vert 0\rangle.
		\end{split}
	\end{equation}
	The transition frequencies between the $\vert0\rangle_{\rm d}$ and $\vert \pm 1\rangle_{\rm d}$ are:
	\begin{equation}\label{omegapm}
		\omega_{\pm} = \frac{\Omega_2}{2}\pm \frac{d_\parallel E_z+3d_\perp E_x}{2}+O(B^2),
	\end{equation}
	which are linear with the electric field but have zero first-order dependence on the magnetic field (Fig.~\ref{Fig-1}(e)). Also, with the suppression of the magnetic field noise, the energy level shift is discerned clearly, which makes it easier to resolve the electric field (Fig.~\ref{Fig-1}(f)).

\begin{figure}
		\centering
		\includegraphics[scale=0.66]{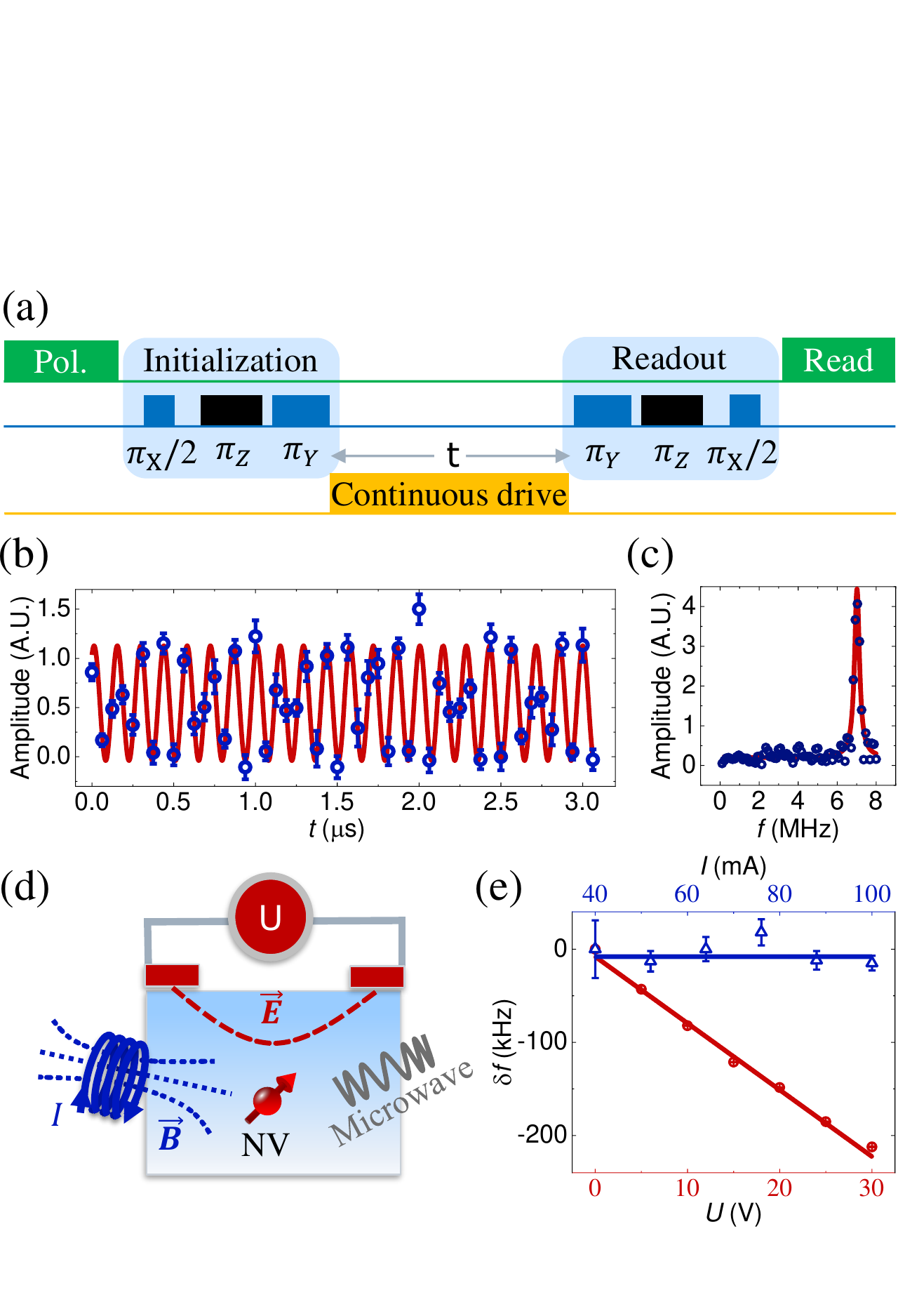}
		\caption{\textbf{Measurement of the electric field.} \textbf{(a)} The pulse sequence. The two green boxes represent the laser polarization and readout of the state of the NV center. The two control pulse groups, "Initialization" and "Readout", each consists of a microwave pulses chain. The blue boxes are normal microwave pulses with different phases, while the black boxes are axial radiofrequency pulses \cite{SOM}. The orange box is the continuous driving pulse with length t. \textbf{(b)} The time-domain signal of the Ramsey experiment. $\Omega_1 = 16$ MHz and $\Omega_2 = 2$ MHz. The circular points are the experimental results with error bars indicating the s.e.m. The red line is a sinusoidal fit. \textbf{(c)} The FFT of the signal of  \textbf{(b)}. The Lorentz fitting gives the electric-field-dependent oscillation frequency $\Omega_1/2-\omega_+$. \textbf{(d)} Sketch of the setup for measuring the electric field. The coil is used to generate the magnetic field by applying current $I$. The electrode is used to provide the electric field. Microwave and radiofrequency pulses are radiated on the NV center by a waveguide. The NV center is roughly 8 $\mu {\rm  m}$ deep from the diamond surface. \textbf{(e)} The frequency shifts with separately varying voltages (red dots) or currents (blue triangles). The frequency shifts show clear linear dependence on the applied voltage, but no observed change on the applied current. All points are experimental results with error bars indicating the s.e.m, and all lines are linear fittings.}
		\label{Fig-2}
	\end{figure}

	We use a Ramsey-like sequence \cite{maze2012free} in the dressed-state space (Fig.~\ref{Fig-2}(a)) to measure the electric field. Thus a superposition of $\vert0\rangle_{\rm d}$ and $\vert+1\rangle_{\rm d}$ or $\vert-1\rangle_{\rm d}$ should be prepared as an initial state. For simplicity, we take $\left(\vert+1\rangle_{\rm d} +\vert0\rangle_{\rm d}\right)/\sqrt{2}$ as an example, which is just $\vert+1\rangle$ in the lab frame. The NV center is first polarized to $\vert0\rangle$ by the laser pulse, and then a chain of microwave pulses are applied to prepare the state into $\vert+1\rangle$. The evolution is given as:
	\begin{equation}\label{UI}
	\vert+1\rangle=U_Y(\pi)U_Z(\pi)U_X(\frac{\pi}2)\vert 0\rangle,
	\end{equation}
	where the $U_Y(\pi)$ and $U_X(\frac{\pi}2)$ are the common $\pi$ and $\pi/2$ pulses with $90^\circ$ phase difference. The $U_Z(\pi)$ is realized by an axial magnetic field pulse \cite{SOM}. Another way for the initialization is introducing a circular polarized microwave $\pi$ pulse with a specially designed microwave radiator \cite{london2014strong}. After the initialization, the continuous driving field is applied, and the NV center evolves in the dressed-state space governed by the Hamiltonian Eq.~\ref{Heff}. During the "free" evolution for a duration of t, the superposition state will accumulate a relative phase between $\vert+1\rangle_{\rm d}$ and $\vert0\rangle_{\rm d}$, and becomes $\left({\rm e}^{i\omega_+ t}\vert+1\rangle_{\rm d} +\vert0\rangle_{\rm d}\right)/\sqrt{2}$. Then the reversed microwave
	pulses chain, marked with "Readout" in Fig.~\ref{Fig-2}(a),  acts as the readout pulse, which transforms the accumulated phase into the population of $\vert0\rangle$. Finally, this population is read out by the photoluminescence \cite{steiner2010universal}.
	
	The time-dependent phase results to an oscillation signal. Due to the existence of an intermediate rotating reference frame with rate $\Omega_1$ \cite{SOM}, this oscillation generally has three frequencies: $\Omega_1/2-\omega_+$, $\Omega_1/2+\omega_+$ and $\Omega_1$. To simplify the measurement, here we use the under-sampling method with the sampling rate of $\Omega_1$. The final normalized under-sampling signal (Fig.~\ref{Fig-2}(c,d)) in the lab frame is given as:
	\begin{equation}\label{S_u}
	S(t)=\frac12(1+\cos((\frac{\Omega_1}2-\omega_+) t)), 
	\end{equation}
	where only one electric-field-dependent frequency remains. 

	In our experiments, we first demonstrate the electrometry with the setup shown in Fig.~\ref{Fig-2}(d). In this setup, microwave and radiofrequency are transmitted by a waveguide to manipulate the NV center, which is embedded in an electronic-grade diamond. A pair of electrodes are electroplated on the diamond surface to generate the electric fields via supplying a  voltage $U$. A coil is placed aside from the diamond to generate the magnetic field via supplying a current $I$. By varying $U$ and $I$ separately, the frequency shifts ($\delta f$) of the under-sampling signal depicted by the Eq.~(\ref{S_u}) are shown in Fig.~\ref{Fig-2}(e). One can see that $\delta f$ shows a clear linear dependence on the voltage $U$, i.e., the electric field. As a contrast, $\delta f$ remains nearly constant with increasing current $I$, ie the magnetic field. The maximum axial magnetic field is roughly 16 $ {\mu}\rm T$, corresponding to an energy shift of $\sim 450$ kHz in the lab frame. Therefore, the electric-field-sensitive and magnetic-field-insensitive properties of the transition between dressed states makes it an ideal candidate for electrometry.
	
	As a practical example, we then use this dressed spin sensor to in situ investigate the noise environment of near-surface NV centers, which is important for quantum sensing \cite{sushkov2014magnetic, grinolds2014subnanometre, shi2015single, muller2014nuclear}. Given that the noise sources on the diamond surface are complicated, including magnetic noise induced by paramagnetic defects \cite{rosskopf2014investigation, romach2015spectroscopy, jamonneau2016competition} and electric noise induced by charge fluctuations \cite{kim2015decoherence, myers2017double, jamonneau2016competition, chrostoski2018electric}, it is essential to deal with them separately, where our method is favorable. We note the double-quantum method can also distinguish the electric and magnetic noise \cite{myers2017double}, but it is not applicable for low-frequency noises (($\leqslant \ $MHz)).
	
	\begin{figure}
		\centering
		\includegraphics[scale=0.66]{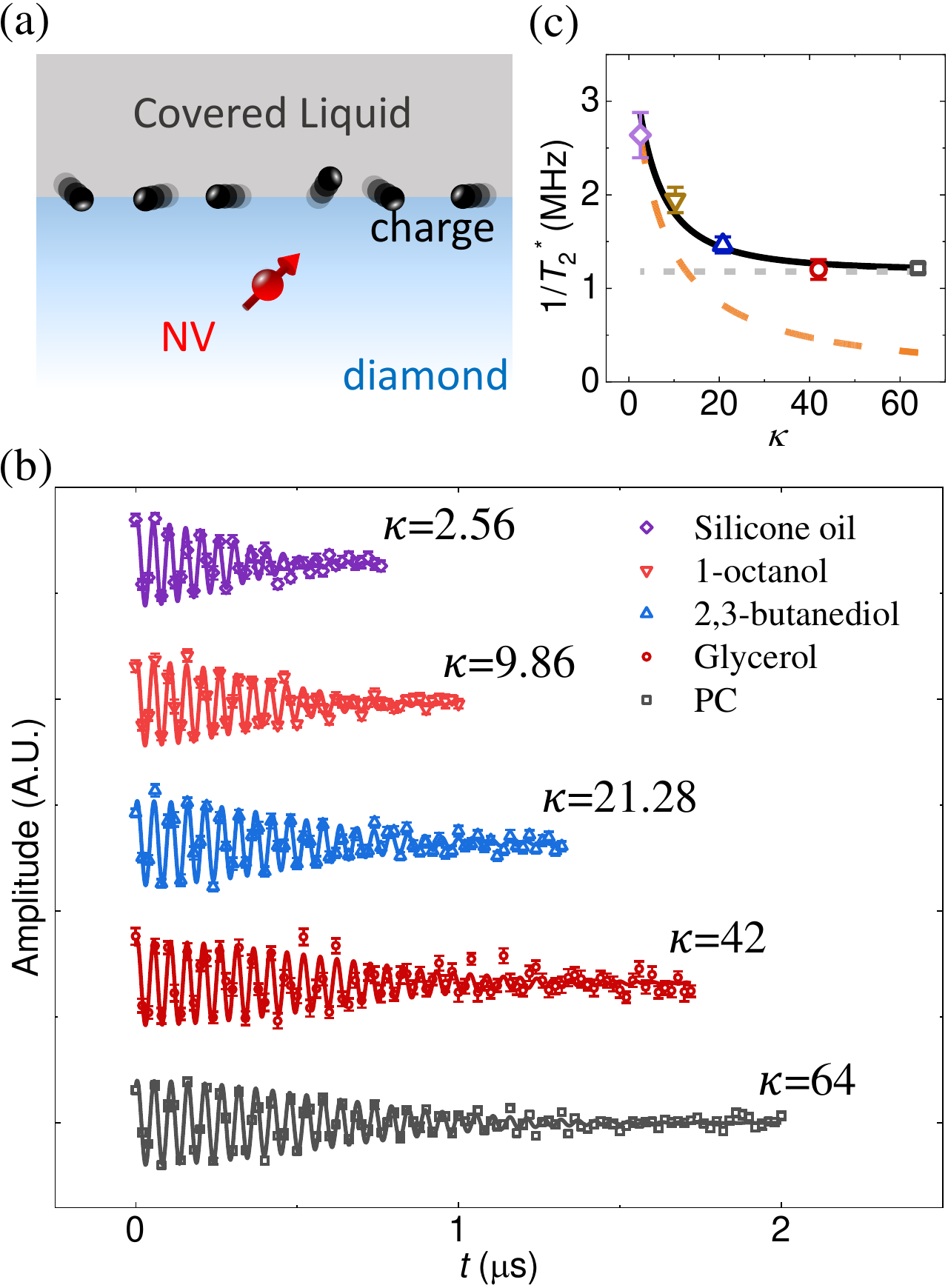}
		\caption{\textbf{Measurement of the electric noise.} \textbf{(a)} The liquid is covered on the diamond surface and the NV center is roughly 8 nm deep from the diamond surface. The motion of the charges on the interface represent the electric noise source. \textbf{(b)} Dephasing process of the NV center with different surface covered liquids measured by the sequence shown in Fig.~\ref{Fig-2}(a). $\Omega_1 = 50$ MHz and $\Omega_2 = 10$ MHz. All points are experimental results, and all lines are sinusoidal decay fits of the form: $\sin(\omega t)*{\rm exp}(-(t/T_2^*)^2)$. Error bars indicate s.e.m.  \textbf{(c)} The fitted dephasing rate $1/{T}_{2}^*$ versus the dielectric permittivity $\kappa$. The experimental results (points with color the same in Fig.~\ref{Fig-3}(b)) show negative correlation with $\kappa$, but diverge from inverse relationship (orange dash line). Error bars indicate the fitting error. The black solid line is the fitting according to Eq.~\ref{T2} with a non-zero noise floor (gray dash line).}
		\label{Fig-3}
	\end{figure}
	Here we measure the dephasing of the near-surface NV centers to evaluate the surface electric noise. A recent experiment found that this noise can be reduced with a surface covered liquid \cite{kim2015decoherence}. To quantitatively investigate this phenomenon, we cover five kind of liquids on the diamond surface separately (Fig.~\ref{Fig-3}(a)): silicone oil (dielectric permittivity $\kappa=2.56$ \cite{tian2001electrorheology}), 1-Octanol ($\kappa=9.86$ \cite{sastry1997dielectric}), 2,3-Butanediol ($\kappa=21.28$ \cite{ghanadzadeh2010dielectric}), Glycerol ($\kappa=42$ \cite{kim2015decoherence}) and propylene carbonate (PC) ($\kappa=64$ \cite{kim2015decoherence}). Fig.~\ref{Fig-3}(b) gives a representative result. With the higher dielectric constant liquid, the lower dephasing rate could be reached. This suggests that the dephasing of the near-surface NV centers are indeed affected by the electric noise on the surface. Presuming the dephasing is solely attributed to the surface electric noise, the dephasing rate $1/{T}_{2}^*$ of the dressed states should be inverse proportion of the dielectric permittivity (Fig.~\ref{Fig-3}(c)) as analyzed below. But we observe an obvious non-zero noise floor in Fig.~\ref{Fig-3}(c), indicating the existence of other surface-irrelevant noises. Since we have suppressed the the magnetic noise and thermal noise to the magnitude one order lower than the floor noise ($\sim 1/T^*_2$) \cite{SOM}, this intrinsic surface-irrelevant noise should still be the electric noise. 
	
	To analysis the origin of the intrinsic electric noise, we use a simple electrostatic model \cite{kim2015decoherence}. The surface electric field is given as:
	\begin{equation}\label{E_s}
	E_{\rm s}\propto\frac1{\kappa_{\rm d}+\kappa_{\rm ext}},
	\end{equation}
	where $\kappa_{\rm d} = 5.7$ and $\kappa_{\rm ext}$ are the dielectric permittivities of the diamond and the surface liquid, respectively.
	The NV center experiences the electric noise arising from two origins, the intrinsic one $\langle E_{\rm i}^2 \rangle$ and the surface one $\langle E_{\rm s}^2 \rangle$. Assuming the electric noise is quasi-static and Gaussian distributed, the dephasing time $T_2^*$ is related as \cite{SOM}:
	\begin{equation}\label{T2}
	\frac1{T_2^*}\propto\sqrt{\langle E_{\rm i}^2 \rangle+(\frac{\kappa_{\rm d}+\kappa_{\rm air}}{\kappa_{\rm d}+\kappa_{\rm ext}})^2\langle E_{\rm s,air}^2 \rangle},
	\end{equation} 
	where $\langle E_{\rm s,air}^2 \rangle$ is the surface electric noise with the diamond surface exposed in air, and $\kappa_{\rm air} = 1$ is the dielectric permittivity of the air.
	\begin{figure}
		\centering
		\includegraphics[scale=0.66]{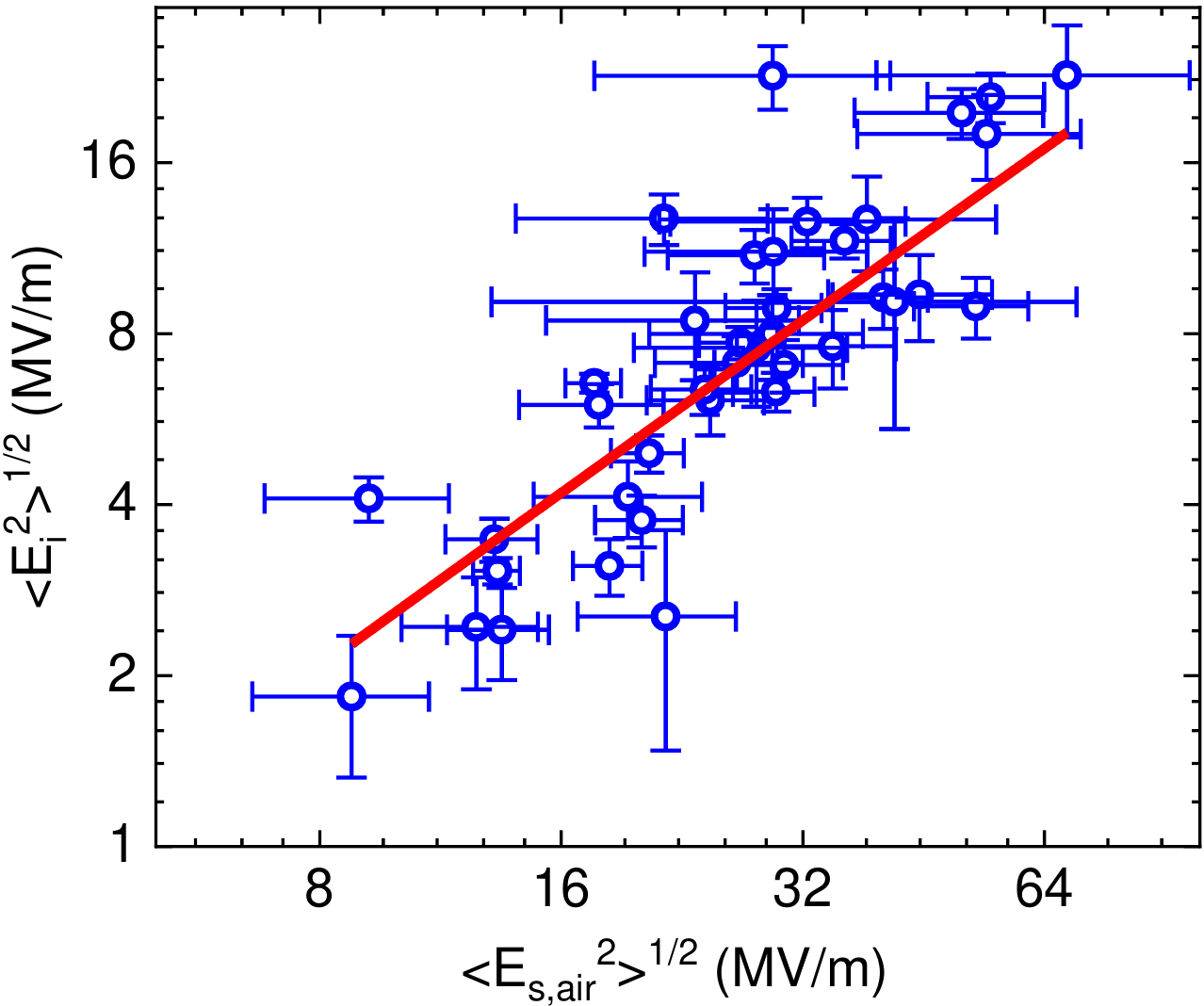}
		\caption{\textbf{Quantitative analysis of the near-surface electric noise.} Relation between of the surface electric noise in air and the intrinsic electric noise. Each point represents a set of measurements similar with Fig.~\ref{Fig-3}. All the data points and error bars are calculated from the fitting parameters. The line is a linear fitting as a guide to the eye.}
		\label{Fig-4}
	\end{figure}
	
	By fitting the data in Fig.~\ref{Fig-3}(c) according to the Eq.~\ref{T2}, we can extract the surface and intrinsic noise. We repeat the measurements on dozens of NV centers, and then get a statistics of $\langle E_{\rm s,air}^2 \rangle^{1/2}$ and $\langle E_{\rm i}^2 \rangle^{1/2}$ as shown in Fig.~\ref{Fig-4}. The result shows that the intrinsic noise is ubiquitous and seems positively correlated to the surface noise, despite a certain margin of error. It was recently reported that there exists an electric field near the diamond surface due to the surface and internal shallow defects \cite{ mittiga2018imaging, bluvstein2019identifying}. We think the instability \cite{broadway2018spatial} or rearrangement \cite{bluvstein2019identifying} of these defects is responsible for the observed electric noise here. The diverseness of the noise arises from the local inhomogeneity of the defects and different NV depths. The latter may attribute to the correlation between internal and surface noise. In addition, the $\langle E^2 \rangle^{1/2}$ is estimated to be on the order of $10^7~\rm{V/m}$ according to the dephasing rate, which consists with the previously work \cite{kim2015decoherence}.

	In conclusion, we have demonstrated a robust method for nanoscale electrometry based on spin sensors in diamond. Our electrometric method is applicable even in the presence of strong magnetic field inhomogeneity or fluctuation, which is favorable for practical applications, for example, the characterization of multiferroic materials \cite{cheong2007multiferroics}. We also use this method to study the noise environment of near-surface NV centers. By excluding the magnetic noise, we observe a quantitative relation between the dephasing rate of NV centers and the dielectric permittivity of surface covered liquids. This finding helps further understanding of the noise environment of near-surface NV centers, which is essential for a wide range of sensing applications, and offers interesting avenues for nanoscale dielectric sensing.
	

	The authors thank Yiheng Lin for helpful discussions. This work was supported by the National Key Research and Development Program of China (Grants No. 2018YFA0306600, 2017YFA0305000 and 2016YFA0502400), the National Natural Science Foundation of China (Grants No. 81788101, 91636217, 11722544, 11761131011, 31971156 and 11775209), the CAS (Grants No. GJJSTD20170001, QYZDY-SSW-SLH004 and YIPA2015370), the Anhui Initiative in Quantum Information Technologies (Grant No. AHY050000), the CEBioM, the national youth talent support program, the Fundamental Research Funds for the Central Universities, the China Postdoctoral Science Foundation (Grant No. BX20180294), and USTC Research Funds of the Double First-Class Initiative (Grant No. YD2340002004).

	\renewcommand\refname{Reference}
	
	\bibliography{citations}

\end{document}


	
	\title{Nanoscale electrometry based on a magnetic-field-resistant spin sensor}
	\author{Rui Li}
	\altaffiliation{These authors contributed equally to this work.}
	\author{Fei Kong}
	\altaffiliation{These authors contributed equally to this work.}
	\author{Pengju Zhao}
	\author{Zhi Cheng}
	\author{Zhuoyang Qin}
	\author{Mengqi Wang}
	\author{Qi Zhang}
	
	\author{Pengfei Wang}
	
	\author{Ya Wang}
	\email{ywustc@ustc.edu.cn}
	\author{Fazhan Shi}
	\email{fzshi@ustc.edu.cn}
	\author{Jiangfeng Du}
	\email{djf@ustc.edu.cn}
	\affiliation{Hefei National Laboratory for Physical Sciences at the Microscale and Department of Modern Physics, University  of  Science  and  Technology of  China (USTC), Hefei, 230026, China}
	\affiliation{CAS Key Laboratory of Microscale Magnetic Resonance, USTC, Hefei, 230026, China}
	\affiliation{Synergetic Innovation Center of Quantum Information and Quantum Physics,\\ USTC, Hefei, 230026, China}
    
	\maketitle

	\section{Samples and setup}\label{SUP0}
		The two electronic-grade diamonds used in this work are both synthesized by chemical
	vapor deposition (CVD) from Element Six. The nitrogen-vacancy (NV) centers in the first diamond are created during the process of the diamond synthesis. An NV center, roughly 8 $\mu $m deep from the diamond surface, is selected to be the electric field sensor. The diamond surface above the NV center is etched to create a microscopic solid-immersion-lenses (SIL) by focused ion beam (FIB) milling to enhance the collection of the photoluminescence \cite{jamali2014microscopic}.
	The second diamond used to investigate the surface noise includes two regions, which are implanted by $^{14}N^+$ with dose density $1\times10^9\ {\rm cm}^{-2}$ at the energy of 5 keV and 70 keV, respectively. Then it is annealed at 1000 $^{\circ}$C to create roughly 8-nm-depth and 85-nm-depth (simulated by the SRIM \cite{ziegler2010srim}) NV centers near the diamond surface \cite{pezzagna2010creation}.

\begin{figure}
		\centering
		\includegraphics[scale=1.2]{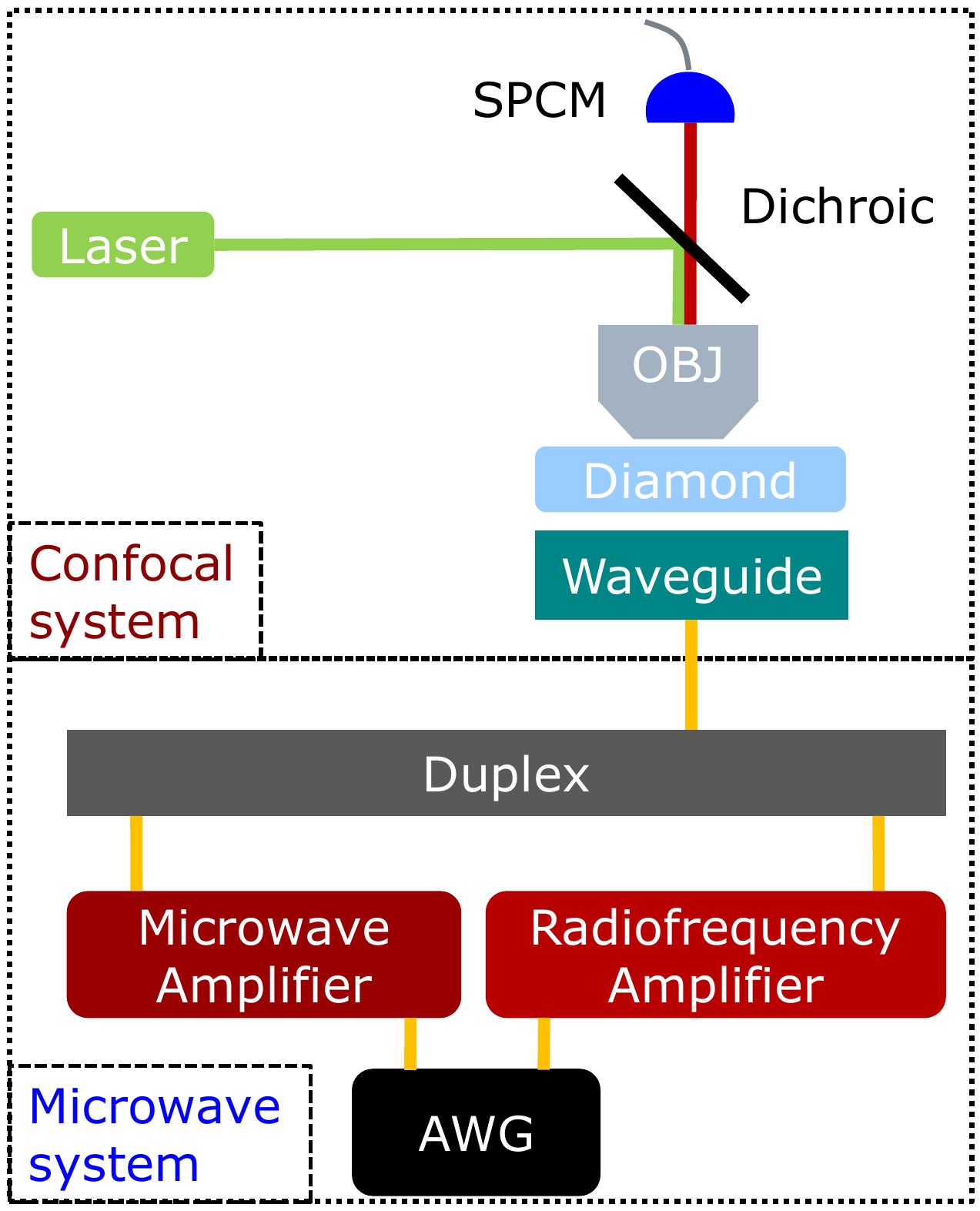}
		\caption{\textbf{Confocal system and Microwave system.} A 532-nm laser and the single photon counting module (SPCM) are used to polarize and read the state of the NV center, separately. The green light is focused and the photoluminescence is collected both by the objective lens. The dichroic mirror is used to separate the two light of different wave lengths. The microwave and radiofrequency signals from the microwave system is transmitted in the waveguide.}
		\label{Fig-S00}
	\end{figure}

\begin{figure}
		\centering
		\includegraphics[scale=1.2]{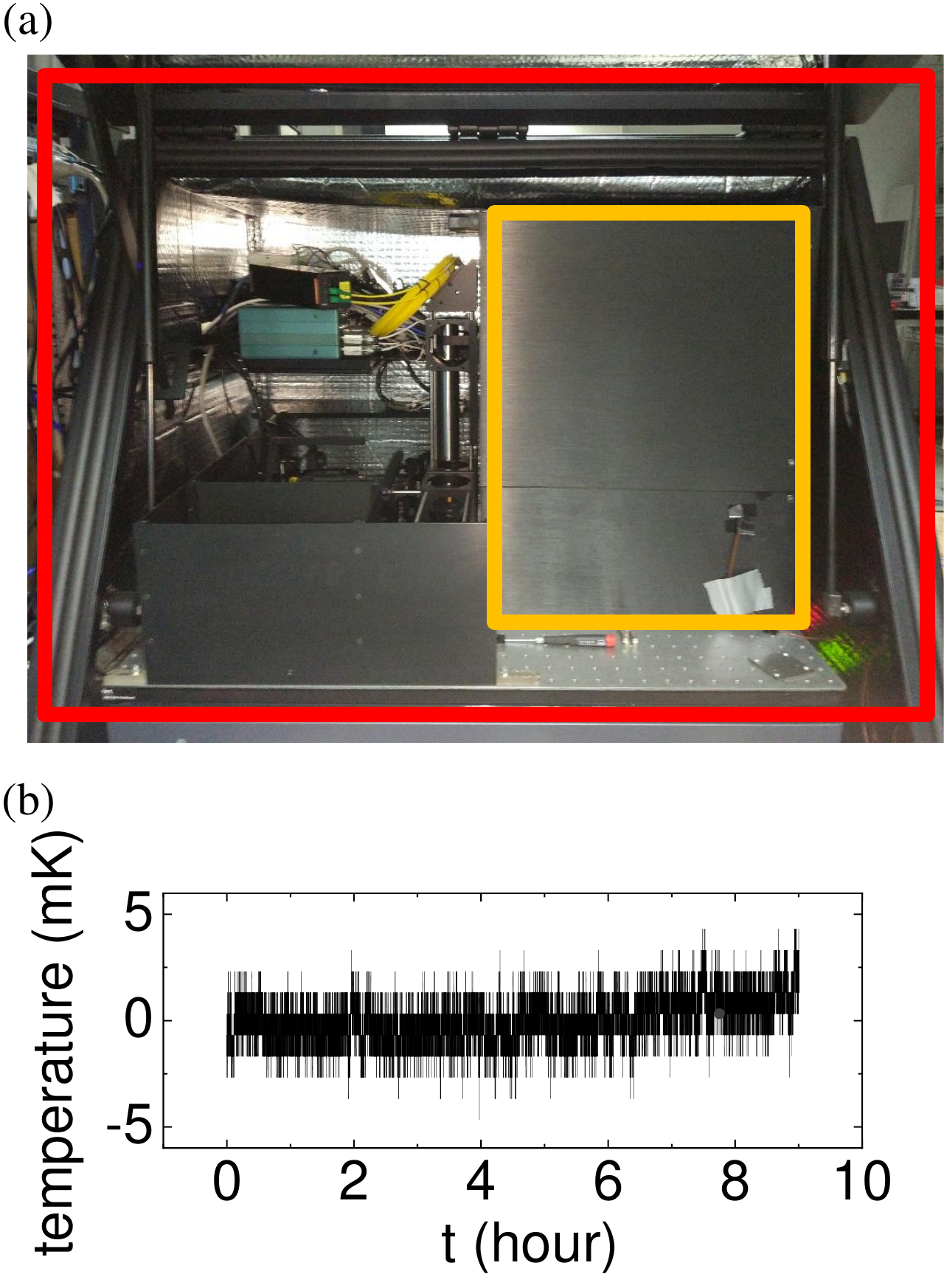}
		\caption{\textbf{Incubators} \textbf{(a)} We use two incubators to control the temperature. The red box (NanoVault) and the yellow box (home-built incubator). \textbf{(b)} The temperature fluctuation in the home-built incubator near the diamond measured in 9 hours.}
		\label{Fig-S0}
	\end{figure}

	The setup is a home-built confocal microscope used to manipulate and read the NV centers. The setup contains three main systems: the optical confocal system, the microwave system and the temperature system. The optical confocal system (Fig.~\ref{Fig-S00}) mainly consists of a fiber-optic laser (CNI, MGL-III-532) and a single photon counting module
	(Perkin Elmer, SPCM-AQRH-14) which are used for polarization and readout of the states of the NV centers. The microwave system (Fig.~\ref{Fig-S00})mainly consists of an arbitrary wave generator (AWG) (Keysight, M8190), a microwave amplifier (Mini-Circuits, ZHL-16W-43+) and a radiofrequency amplifier (Mini-Circuits, LZY-22+). The AWG generates the microwave and the radiofrequency signals from two separate ports. The two ports are connected with the two amplifiers respectively, and then jointed by a duplex (Marki, DPXN-1). The signal from the duplex is finally delivered by a waveguide to manipulate the NV centers. The temperature system contains two incubators (Fig.~\ref{Fig-S0}(a)). The larger one is obtained commercially (Herzan, NanoVault) with a commercial temperature controller (WATLOW, EZ-ZONE), where the temperature fluctuation is controlled within $\pm\ 0.1$ K. The diamond is placed in a home-built incubator which contained in the larger incubator (Fig.~\ref{Fig-S0}(a)). The temperature controller of this incubator is obtained commercially (Standford Instrument, PTC10). In the home-built incubator and near the diamond, the temperature is controlled $2\sim3$ K higher than the room temperature and the fluctuation is within 10 mK (Fig.~\ref{Fig-S0}(b)).
	\section{Hamiltonian in the dressed frame}\label{SUP1}
	The Hamiltonian of an NV center in the lab frame (energy level shown in Fig. 1(a) of the main text) is described as:
	\begin{equation}\label{H0}
	H_0=(D+d_\parallel E_z)S_z^2 + \gamma \vec{B}\cdot\vec{S} -d_\perp E_x (S_x^2-S_y^2)+d_\perp E_y (S_x S_y+S_y S_x),
	\end{equation}
	where $D$ is the zero field split of the NV center, $\gamma$ is the gyromagnetic ratio of the electron spin, $\vec{S}$ is the electron spin operator, $d_\parallel$ and $d_\perp$ are the axial and non-axial electric dipole moments, $\vec{B}$ and $\vec{E}$ are the magnetic and electric field, separately.
	The Hamiltonian of a continuous phase-modulated microwave driving field on the NV center is given as:
	\begin{equation}\label{HPMMW}
	H_1=\Omega_1\cos(f t+\frac{2\Omega_2}{\Omega_1}\sin(\Omega_1 t))S_x.
	\end{equation}
	$\Omega_1$ is the Rabi frequency between the bare states, $m_s=0$ ($\vert 0 \rangle$ henceforth) and degenerate $m_s=\pm 1$ ($\vert \pm 1\rangle$ henceforth). $f$ is the base frequency of the continuous microwave. $\Omega_2$ here is arranged to control the magnitude of the phase modulation.
	The total Hamiltonian of the NV center is:
	\begin{equation}\label{H_total}
	H=H_0+H_1.
	\end{equation}
	First, if we set the frequency $f=D$ and apply the transformation 
	$ U_1= {\rm exp}  [i(D t+(2\Omega_2/\Omega_1)\sin(\Omega_1 t))S_z^2]$ onto the Hamiltonian (Eq.~\ref{H_total}). With the rotating wave approximation ($D\gg\Omega_1$ and $D\gg B$ are required), the Hamiltonian becomes:
	\begin{equation}\label{HR1}
		\begin{split}
	H_{R1}&=U_1HU_1^{-1}-iU_1\frac{{\rm d}U_1^{-1}}{{{\rm d}t}} \\
	&= d_\parallel E_z S_z^2 + \gamma B_z S_z -d_\perp E_x (S_x^2-S_y^2)+d_\perp E_y (S_x S_y+S_y S_x) + \frac{\Omega_1}2 S_x - 2\Omega_2 \cos(\Omega_1 t)S_z^2.
		\end{split}
	\end{equation}
	If $\Omega_1$ is controlled an order larger than $\gamma B_z, d_\perp E_x, d_\perp E_y, d_\parallel E_z$, then with the perturbation theory, the Hamiltonian (Eq.~\ref{HR1}) is approximated as:
	\begin{equation}\label{HR1A}
	H_{R1}^{'}=(\frac{\Omega_1}2+\Delta) S_x - (\frac12d_\parallel E_z+\frac32 d_\perp E_x) S_x^2- \Omega_2 \cos(\Omega_1 t)(S_z^2-S_y^2),
	\end{equation}
	where $\Delta$ is defined as:
	\begin{equation}\label{Delta}
	\Delta=\frac{(\gamma B_z)^2}{\Omega_1}+\frac{(d_\parallel E_z-d_\perp E_x)^2}{4\Omega_1}+\frac{(d_\perp E_y)^2}{\Omega_1}.
	\end{equation}
	Then by applying another transformation $U_2 = {\rm exp}[i(\frac{\Omega_1}2 t)S_x] $ and with the rotating wave approximation ($\Omega_1\gg\Omega_2$ is required), the Hamiltonian becomes:
	\begin{equation}\label{HR2}
	H_{R2}=U_2 H_{R1}^{'} U_2^{-1}-iU_2\frac{{\rm d}U_2^{-1}}{{{\rm d}t}} =(\frac{\Omega_2}2 + \frac{\Delta^2}{\Omega_2})(S_z^2-S_y^2) +(\frac12 d_\parallel E_z +\frac32 d_\perp E_x)S_x^2.
	\end{equation}
	The eigenstates of the Hamiltonian (Eq.~\ref{HR2}) are $(\vert+1\rangle+\vert-1\rangle)/\sqrt2$, $(\vert+1\rangle-\vert-1\rangle)/\sqrt2$ and $\vert0\rangle$, which are defined as $\vert+1\rangle_{\rm d}$, $\vert0\rangle_{\rm d}$ and $\vert-1\rangle_{\rm d}$ correspondingly. The subscript 'd' represents the states in the dressed frame. And in the dressed frame, the Hamiltonian (Eq.~\ref{HR2}) is given as:
	\begin{equation}\label{Hd}
	H_{\rm d}=(\frac{\Omega_2}2 + \frac{\Delta^2}{\Omega_2})S_{z,{\rm d}} + (\frac12 d_\parallel E_z +\frac32 d_\perp E_x)S_{z,{\rm d}}^2.
	\end{equation}
	Set $\Omega_2 \gg \Delta$, we get the approximated Hamiltonian in the dressed frame:
	\begin{equation}\label{HdA}
	H_{\rm dA}=\frac{\Omega_2}2 S_{z,{\rm d}}  + (\frac12 d_\parallel E_z+\frac32 d_\perp E_x)S_{z,{\rm d}}^2.
	\end{equation}
	In the dressed frame, $\Omega_2$ controls the energy transition frequencies while the frequency shift is linear with the electric field.
	\section{$U_Z(\pi)$ pulse implementation}\label{SUP2}
\begin{figure}
		\centering
		\includegraphics[scale=1.2]{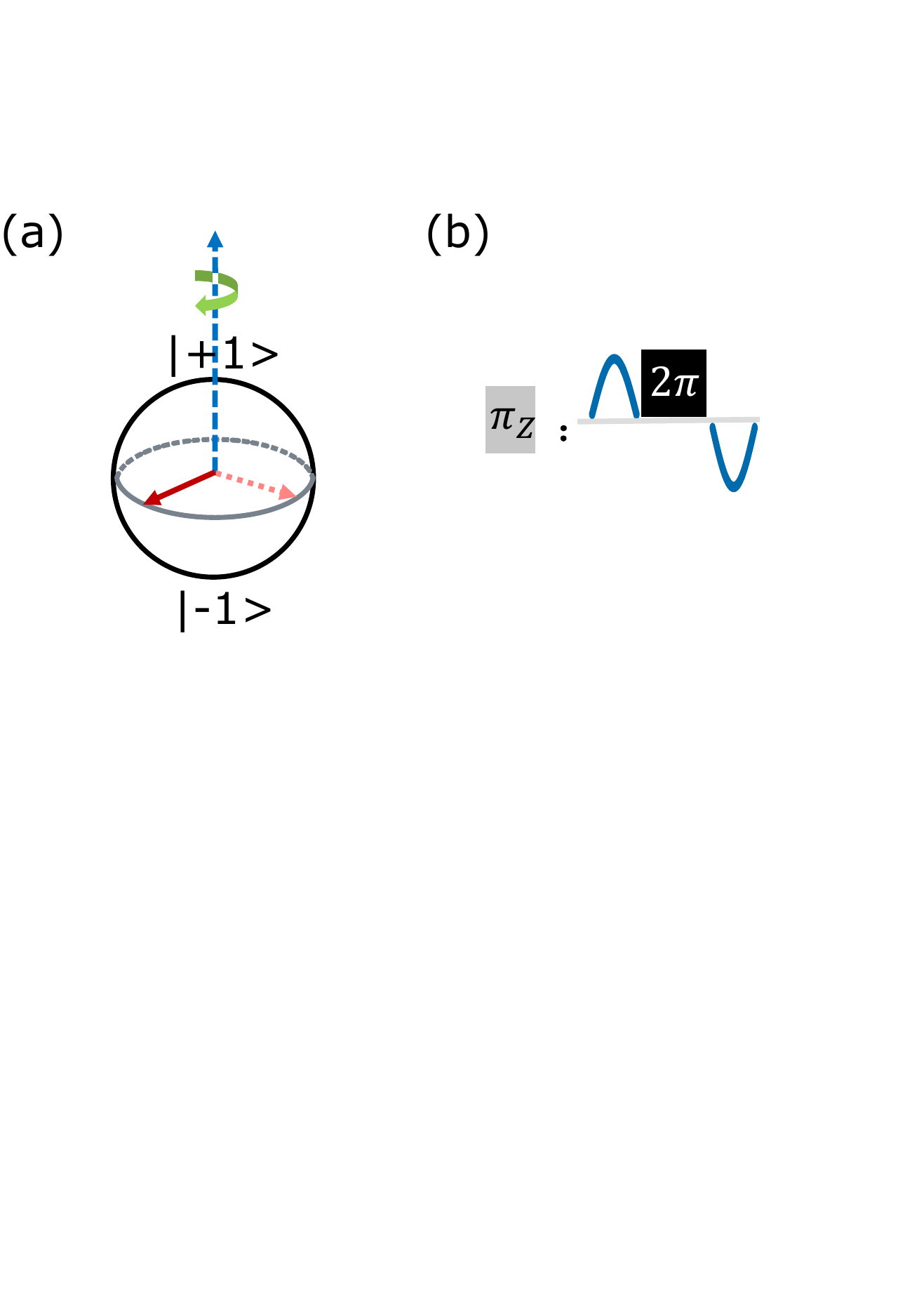}
		\caption{\textbf{$\bm {U_Z(\pi)}$ pulse.} \textbf{(a)} When an axial bias pulse is applied on the NV center, the Bloch vector would rotate along the z axis. With the control of the bias pulse duration, a $\pi$ phase shift could be reached between $\vert+1\rangle$ and $\vert-1\rangle$ states. \textbf{(b)} The $U_Z(\pi)$ pulse chain we used consists of two radiofrequency pulses and a $2\pi$ microwave pulse along X axis.}
		\label{Fig-S1}
	\end{figure}

	The trivial pulse of $U_Z(\pi)$ is implemented by applying an axial bias radiofrequency pulse which results in a $\pi$ phase accumulated between $\vert+1\rangle$ and $\vert-1\rangle$ states (shown in Fig~\ref{Fig-S1}(a)). However, a bias pulse into an amplifier would distort the waveform and decrease the $U_Z(\pi)$ fidelity. So we use a pulse chain consists of two radiofrequency pulses and a microwave pulse (shown in Fig~\ref{Fig-S1}(b)):
	\begin{equation}\label{Uz}
	U_Z(\pi)=U_{Z,{\rm real}}(\pi/2)U_X(2\pi)U_{Z,{\rm real}}(-\pi/2).
	\end{equation}
	It should be noted that the $U_Z(\pi)$ of Eq.~\ref{Uz} is not the real axial $\pi$ pulse (denoted as $U_{Z,{\rm real}}(\pi)$) that only causes a phase shift between $\vert+1\rangle$ and $\vert-1\rangle$ states. It also swaps the population of those two states because of $U_X(2\pi)$. The difference would change the read state, which could be revised by the order of the readout pulse chain mentioned in the main text. For example, the initial pulse chain with $U_Z(\pi)$ prepares the state:
	\begin{equation}\label{UI}
	\vert+1\rangle=U_Y(\pi)U_Z(\pi)U_X(\frac{\pi}2)\vert 0\rangle,
	\end{equation}
	while the initial pulse chain with $U_{Z,{\rm real}}(\pi)$ would prepare the state:
	\begin{equation}\label{UI2}
	-\vert+1\rangle=U_Y(\pi)U_{Z,{\rm real}}(\pi)U_X(\frac{\pi}2)\vert 0\rangle,
	\end{equation}
	where the overall phase could be ignored. While in the read pulse chain,
	\begin{equation}\label{UI3}
	\left|\langle 0 \vert U_X(\frac{\pi}2)U_Z(\pi)U_Y(\pi)  \vert \Phi \rangle \right|^2=\left|\langle 0 \vert U_Y(\frac{\pi}2)U_{Z,{\rm real}}(\pi)U_X(\pi) \vert \Phi \rangle \right|^2,
	\end{equation} 
	where $ \vert\Phi \rangle$ is the state after "free" evolution mentioned in the main text.
	In conclusion, the difference between $U_Z(\pi)$ and $U_{Z,{\rm real}}(\pi)$ would not affect the main results of this work.
	
	\section{Evolution in the dressed frame}\label{SUP3}
\begin{figure}
		\centering
		\includegraphics[scale=1.2]{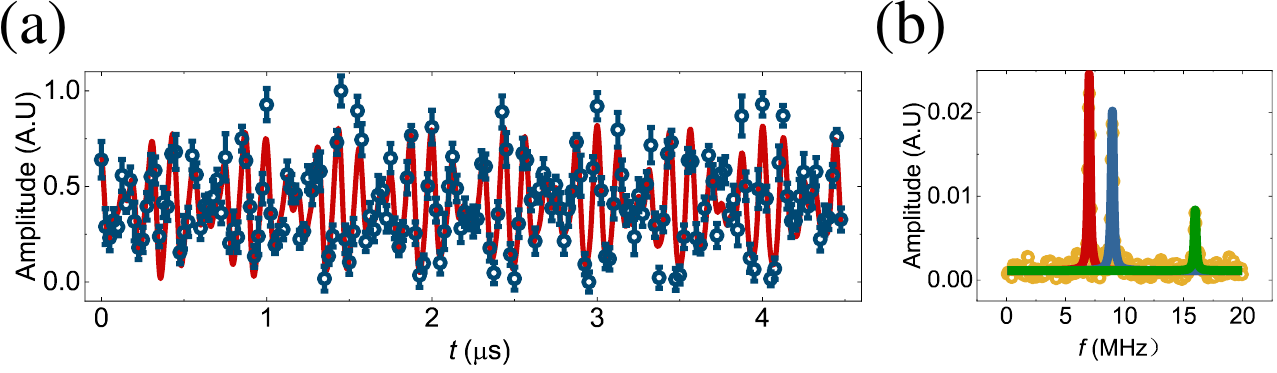}
		\caption{\textbf{Signal.} \textbf{(a)} The result measured by the sequence shown in Fig. 2(b) in the main text. $\Omega_1 = 16$ MHz and $\Omega_2 = 2$ MHz. It is fitted in the red line according to the Eq.~\ref{S}. \textbf{(b)} The FFT of the signal of  \textbf{(a)}. There are three peaks which are fitted by Lorentz shape lines respectively. The first two fitted peaks (in red and blue) would shift linear with the applied voltage while the last and lower peak (in green) is constant.}
		\label{Fig-S2}
	\end{figure}

	Evolution under the sequence (shown in Fig. 2(b) in the main text) is guided by the Hamiltonian (Eq.~\ref{HdA}).  If we take $\left(\vert+1\rangle_{\rm d} +\vert0\rangle_{\rm d}\right)/\sqrt{2}$ as an initial state the same in the main text, the signal is given as:
	\begin{equation}\label{S}
		S(t)=\frac18(2\cos(\frac{\Omega_1+\Omega_2+3d_\perp E_x+d_\parallel E_z}2 t)+2\cos(\frac{\Omega_1-\Omega_2-3d_\perp E_x-d_\parallel E_z}2 t)+3+\cos(\Omega_1 t)).
	\end{equation}
	If we set the applied voltage $U$, i.e., $E_x$ and $E_z$, equal to zero, the result is shown in Fig.~\ref{Fig-S2}(a). The three fitted frequencies in Fig.~\ref{Fig-S2}(b) are $(\Omega_1-\Omega_2)/2$, $(\Omega_1+\Omega_2)/2$ and $\Omega_1$ respectively, of which the amplitudes are roughly $2:2:1$.
	If we set the sampling rate to $\Omega_1$, i.e., $t = 2\pi n/\Omega_1=nT$ with $n = 0,1,2,...$, which is an under-sampling, the signal is given as:
	\begin{equation}\label{S_u}
	S_u(n)=\frac12(1+ \cos(\frac{\Omega_1-\Omega_2-3d_\perp E_x-d_\parallel E_z}2 nT)).
	\end{equation}
	
	\section{Dephasing time $T_2^*$ in the dressed frame}\label{SUP4}
	Assuming the magnitude of the electric field from the noise source is Gaussian distributed and the direction is uncorrelated, the under-sampling signal is given as:
	\begin{equation}\label{S_u_decay1}
	\begin{split}
	S_u(t)&=(\frac1{\sqrt{2\pi}\sigma})^3\int_{-\infty}^{\infty}dE_x\int_{-\infty}^{\infty}dE_y\int_{-\infty}^{\infty}dE_z\frac12(1+\cos(\frac{\Omega_1-\Omega_2-3d_\perp E_x-d_\parallel E_z }2 t)){\rm exp}{(-\frac{E_x^2+E_y^2+E_z^2}{2\sigma^2})} \\
	&=\frac12(1+\cos(\frac{\Omega_1-\Omega_2}2 t){\rm exp}(-\frac{d_{\rm eff}^2\sigma^2 t^2}{2})) = \frac12(1+\cos(\frac{\Omega_1-\Omega_2}2 t){\rm exp}(-\frac{t^2}{T_2^{*2}})), 
	\end{split}
	\end{equation}
	where $d_{\rm eff} = \sqrt{((3 d_\perp)/2)^2 + (d_\parallel/2)^2}$ is the effective dipole moment. The $ \langle E^2 \rangle^{1/2} \propto \sigma$ is the noise magnitude of the electric field. 
	As a result, $1/T_2^* \propto \langle E^2 \rangle^{1/2}$. 
	Assuming the dephasing is solely attributed to the surface noise and 
	using the simple electrostatic model mentioned in the main text, $E\propto1/(\kappa_{\rm d}+\kappa_{\rm ext})$, we could get 
	\begin{equation}\label{Tkappa}
	1/T_2^{*}\propto \frac1{\kappa_{\rm d}+\kappa_{\rm ext}}.
	\end{equation}
	If considering the intrinsic electric noise and assuming there is no correlation between the surface electric field and the intrinsic one, we could get $\langle E^2 \rangle = \langle E_{\rm i}^2 \rangle+\langle E_{\rm s}^2 \rangle$.
	Furthermore, while the liquid covers on the diamond surface, the surface electric noise is given as:
	\begin{equation}\label{Es}
	\langle E_{\rm s}^2 \rangle = \frac{(\kappa_{\rm d}+\kappa_{\rm air})^2}{(\kappa_{\rm d}+\kappa_{\rm ext})^2} \langle E_{\rm s,air}^2 \rangle.
	\end{equation} 
	So we could get:
	\begin{equation}\label{TEi}
	\frac1{T_2^*}\propto\sqrt{\langle E_{\rm i}^2 \rangle+(\frac{\kappa_{\rm d}+\kappa_{\rm air}}{\kappa_{\rm d}+\kappa_{\rm ext}})^2\langle E_{\rm s,air}^2 \rangle},
	\end{equation}
	where the parameters are defined the same in the main text.
	
	\section{Estimate the noise sources}\label{SUP5}
	If the fluctuation of the zero field split ($\delta D$), which comes from the thermal fluctuation \cite{acosta2010temperature},  and the fluctuation of the Rabi frequency ($\delta \Omega_1$), which comes from the unstable of the microwave circuits and amplifier, are both considered, the Hamiltonian Eq.~\ref{H_total} approximated at least second order is given as:
	\begin{equation}\label{HdC}
	H_{\rm d}^{'}=(\frac{\Omega_2}2 + \frac{\Delta_2^2}{\Omega_2})S_{z,{\rm d}} +(\frac12\delta D + \frac12 d_\parallel E_z +\frac32 d_\perp E_x)S_{z,{\rm d}}^2,
	\end{equation}
	where $\Delta_2$ is defined as:
	\begin{equation}\label{Delta2}
	\Delta_2=\frac{\delta \Omega_1}2+\frac{(\gamma B_z)^2}{\Omega_1}+\frac{(\delta D+d_\parallel E_z-d_\perp E_x)^2}{4\Omega_1}+\frac{(d_\perp E_y)^2}{\Omega_1}.
	\end{equation}
	$\Omega_2$ is the phase modulation amplitude, which could be controlled precisely with an arbitrary wave generator (AWG). The phase noise of the equipment is usually low enough to be ignored. As a result, except the linear electric terms, there are two other terms, $\Delta_2$ and $\delta D$, would influence the dephasing of the dressed states of the NV center. To convince one that the electric noise dominates the dephasing between the $\vert+1\rangle_{\rm d}$ and $\vert 0 \rangle_{\rm d}$ measured in the main text, we demonstrate that the noises come from the other two terms could be neglected.
	
	\subsection{Noise comes from $\Delta_2$}\label{SUP3_A}
\begin{figure}
		\centering
		\includegraphics[scale=1.2]{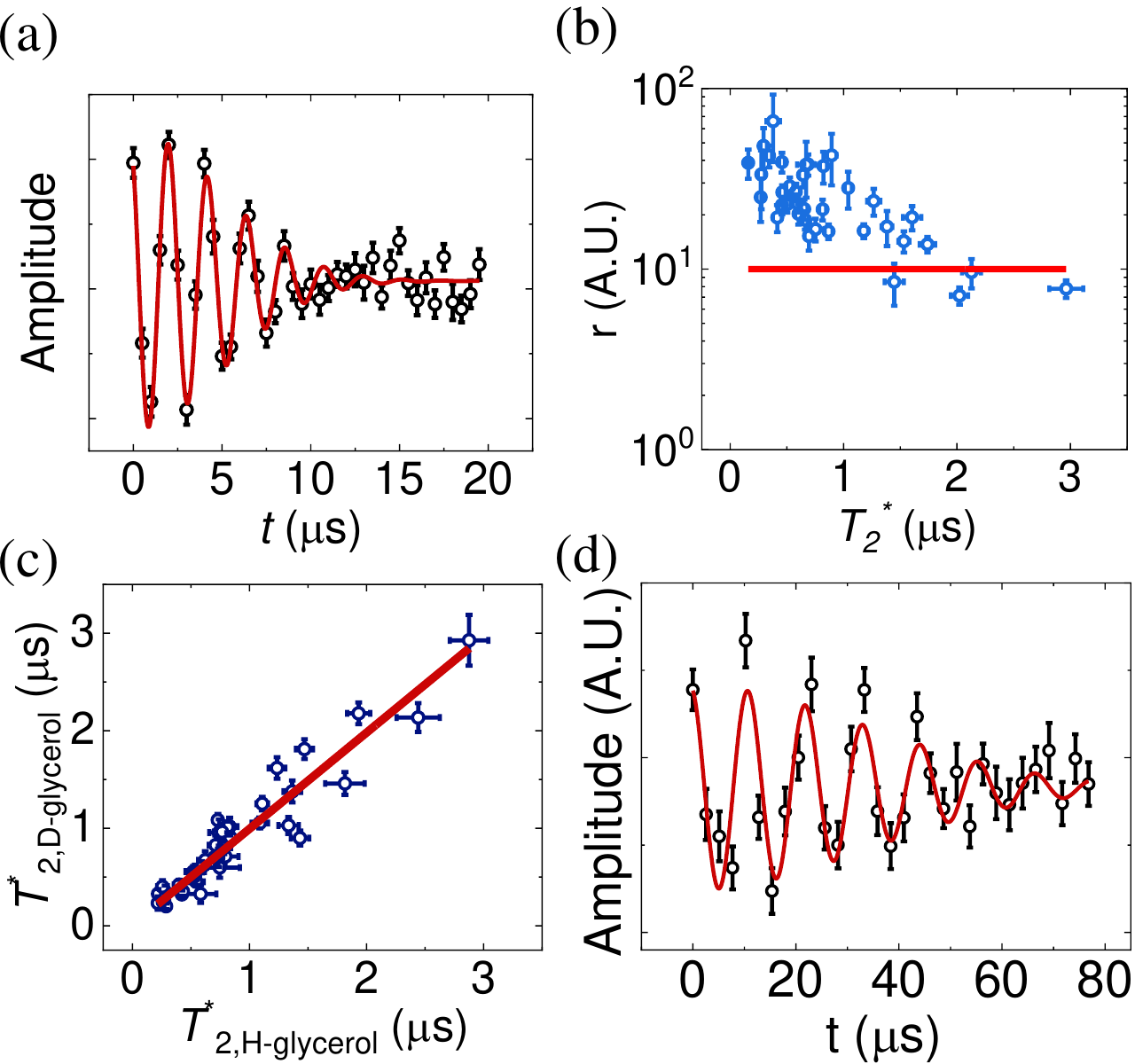}
		\caption{\textbf{Investigation of noise sources.} \textbf{(a)} The dephasing caused by the term $\Delta_2^2/\Omega_2$ between the $\vert+1\rangle_{\rm d}$ and $\vert-1\rangle_{\rm d}$ of one of the measured NV centers. $\Omega_1 = 50$ MHz and $\Omega_2 = 10$ MHz. \textbf{(b)} The ratio of the noise defined in Eq.~\ref{nr}. The horizontal axis is the dephasing time between the $\vert+1\rangle_{\rm d}$ and $\vert0\rangle_{\rm d}$ of the NV center when the diamond surface is covered with PC. Each point represents a measured NV centers. The result shows that the noise magnitude between the $\vert+1\rangle_{\rm d}$ and $\vert0\rangle_{\rm d}$ is more than one order than that between the $\vert+1\rangle_{\rm d}$ and $\vert-1\rangle_{\rm d}$ for most measured NV centers. The red line is at 10 as a reference. \textbf{(c)} The relation of the dephasing time between the $\vert+1\rangle_{\rm d}$ and $\vert 0 \rangle_{\rm d}$ when the diamond surface is covered by H-glycerol and D-glycerol, separately. The data points (blue circle) are linear fitted with a red line. \textbf{(d)} The dephasing between the $\vert+1\rangle_{\rm d}$ and $\vert0\rangle_{\rm d}$ of a roughly 85-nm-depth NV center, of which the dephasing time is around 50 $\mu$s. $\Omega_1 = 50$ MHz and $\Omega_2 = 10$ MHz.}
		\label{Fig-S3}
	\end{figure}

	In the dressed states, we measure the dephasing between the $\vert+1\rangle_{\rm d}$ and $\vert-1\rangle_{\rm d}$ of which the noise only comes from the term $\Delta_2^2/\Omega_2$ (Fig.~\ref{Fig-S3}(a)). We measure dozens of NV centers the same as the NV centers in the main text. We compare it to the dephasing rate between the $\vert+1\rangle_{\rm d}$ and $\vert0 \rangle_{\rm d}$ while the diamond surface is covered by the propylene carbonate (PC) ($\kappa=64$ \cite{kim2015decoherence}). It should be noted that the noise from the term $\Delta_2^2/\Omega_2$ contribute twice to the dephasing between the $\vert+1\rangle_{\rm d}$ and $\vert-1\rangle_{\rm d}$ compared to the dephasing between the $\vert+1\rangle_{\rm d}$ and $\vert0\rangle_{\rm d}$. So we defined the noise ratio:
	\begin{equation}\label{nr}
	r=(1/T_{2, (\vert+1\rangle ,\vert 0\rangle)}^*)/(1/2/T_{2,(\vert+1\rangle ,\vert -1\rangle)} ^*),
	\end{equation}
	The results are shown in Fig.~\ref{Fig-S3}(b). Even at this highest dielectric permittivity liquid we used, i.e., the lowest electric noise, one order of magnitude of the noise ratio for almost all of the NV center indicates the noise coming from $\Delta_2$ could be neglected compared to the measured noise in the main text.
	
	Besides, we compare the dephasing time measured when H-glycerol or D-glycerol is covered on the diamond surface, separately (Fig.~\ref{Fig-S3}(c)). As the magnetic noise in $\Delta_2$ has been proved to be suppressed already, we observe a linear relation with fitted slope of $0.99\pm0.06$ between the two sets of dephasing time, which is different from the observation in Ref.~\cite{kim2015decoherence}.  
	
	\subsection{Noise comes from $\delta D$}\label{SUP3_B}
	The experiments are operated in an incubator where the temperature fluctuation is controlled within 10 mK (Fig~.\ref{Fig-S0}(a)) to suppress the noise comes from the term $\delta D$ \cite{acosta2010temperature}. Furthermore, we measure the dephasing between the $\vert+1\rangle_{\rm d}$ and $\vert0\rangle_{\rm d}$ of a roughly 85-nm-depth NV center with the same parameters used in the main text, e.g., $\Omega_1 = 50$ MHz and $\Omega_2 = 10$ MHz, except the sampling rates and the measuring time(Fig.~\ref{Fig-S3}(d)). Assuming the dephasing is totally attributed to the term $\delta D$, the dephasing rate is at most 20 kHz. Compared to the noise floor (at least $\sim 200\ {\rm kHz}$, calculated by the data shown in Fig .4(a) in the main text) measured in the main text, the noise comes from the temperature fluctuation could be neglected.
	
	As a result, the two terms, $\Delta_2$ and $\delta D$, cannot cause such noise floor shown in Fig .3(c) in the main text. The noise floor of the dephasing rates between the dressed states of the NV center, $\vert+1\rangle_{\rm d}$ and $\vert0\rangle_{\rm d}$, mainly comes from the electric noise. Therefore, the method is proved to be a robust and magnetic-field-resistant electric field metrology.

	\renewcommand\refname{Reference}
	
	\bibliography{citations}